\documentstyle{article}
\newcommand{\be}{\begin{equation}}
\newcommand{\ee}{\end{equation}}

\begin{document}
\title{Creating quanta with `annihilation' operator}
\author{S.S. Mizrahi
\thanks{E-mail: salomon@df.ufscar.br}
and V.V. Dodonov
\thanks{E-mail: vdodonov@df.ufscar.br}
\thanks{on leave from Lebedev Physical Institute and Moscow Institute of
Physics and Technology, Russia}
\\
Departamento de F\'\i sica, CCT, Universidade Federal de S\~ao Carlos,\\
Rod. Washington Luiz Km 235, S\~ao Carlos, 13565-905, SP, Brazil}

\date{}

\maketitle

\begin{abstract}
An asymmetric nature of the boson `destruction' operator $\hat{a}$
and its `creation' partner $\hat{a}^{\dagger}$ is made apparent by
applying them to a quantum state $|\psi \rangle $ different from the
Fock state $|n\rangle $. We show that it is possible to {\em increase} (by
many times or by any quantity) the mean number of quanta in the new
`photon-subtracted' state $\hat{a}|\psi \rangle $.
Moreover, for certain `hyper-Poissonian' states $|\psi \rangle $
the mean number of quanta in the (normalized)
state $\hat{a}|\psi \rangle $ can be much greater than in the `photon-added'
state $\hat{a}^{\dagger}|\psi \rangle $.
The explanation of this `paradox' is given and some examples
elucidating the meaning of Mandel's $q$-parameter and the
exponential phase operators are considered.
\end{abstract}


\section{Introduction}

The non-Hermitian bosonic operators $\hat{a}$ and $\hat{a}^{\dagger}$
of the harmonic oscillator,  satisfying the canonical
commutation relation
$[\hat{a},\hat{a}^{\dagger}]=1$, are usually called `annihilation'
\cite{annih} (or `destruction' \cite{destr,sakurai}) and `creation'
operators (perhaps, only P A M Dirac, in his book \cite{Dirac}, did
not use these terms, introducing instead of $\hat{a}$ and
$\hat{a}^{\dagger}$ the `complex dynamical variables' $\bar{\eta}$ and
$\eta$.) This is due to their action on the Fock (number) state $|n
\rangle$ (eigenstate of the number operator $\hat{n}= \hat{a}^{\dagger}
\hat{a}$) \cite{Fock}
\be \hat{a}\left| n\right\rangle =\sqrt{n}\,| n-1\rangle , \label{an}
\ee
\be \hat{a}^{\dagger}| n\rangle =\sqrt{n+1}\,| n+1\rangle .
\label{adagn} \ee
Therefore, there is some belief that the operator $\hat{a}$ may
`destruct' quanta (photons) in an {\em arbitrary\/} state $|\psi
\rangle $. This belief is reflected even in the name
`photon-sub\-tracted state', sometimes used for the state
$\hat{a}|\psi \rangle $ \cite{Dakna,LuH,Wang00}. However, this
concept, guided maybe more by intuition than by a sound proof, seems
to be misleading when one deals not with single Fock states, but with
their superpositions or quantum mixtures.

The better known counter-example is the harmonic
oscillator coherent state $|\alpha \rangle$:
since it is an eigenstate of $\hat{a}$,
the state $\hat{a}|\alpha \rangle / |\alpha|$ has the same mean
number of photons as $|\alpha \rangle$ (see a similar remark in \cite{LuH}).
But as we will see below, this example is not the only one.
In contradistinction, when
operator $\hat{a}^{\dagger}$ is applied on an arbitrary state
$|\psi\rangle $ it produces a new one, whose mean number of quanta is always
greater than that of $|\psi \rangle $, therefore justifying its
designation as `creation operator'.
Our aim is to provide a deeper investigation of the problem
of `photon subtraction' or `photon addition' for {\em arbitrary\/}
quantum states, drawing readers'
attention to apparent `paradoxes' and giving possible explanations.

The plan of the paper is as follows. In Section 2 we discuss the
relations between the `photon excess' in the `photon-subtracted' states
and Mandel's parameter, illustrating them in the example of superposition
of two Fock states. The concept of `hyper-Poissonian' states is
introduced in Section 3. The examples of such states include certain
superpositions of the coherent and vacuum states and the negative
binomial states. New families of `logarithmic' states corresponding
to the given values of the mean number of quanta and Mandel's parameter
are considered in Section 4. A distinguished role of the exponential
phase operators and their eigenstates -- the phase coherent states --
is discussed in Section 5. Section 6 concludes the paper.

\section{`Photon-subtracted' states, `photon excess' and Mandel's parameter}
For any normalized state
\be | \psi \rangle = \sum_{n=0}^{\infty }c_{n}| n\rangle, \qquad
\sum_{n=0}^{\infty }\left| c_{n}\right| ^{2} =1, \label{psiin}
\ee%
the normalized `photon-subtracted state' is given by
\begin{equation}
\left| \psi_{-}\right\rangle = \frac{1}{\sqrt{\bar{n}}}\hat{a}| \psi
\rangle = \frac{1}{\sqrt{\bar{n}}} \sum_{n=1}^{\infty
}c_{n}\sqrt{n}\,| n-1\rangle
=\frac{1}{\sqrt{\bar{n}}}\sum_{n=0}^{\infty }c_{n+1}\sqrt{n+1}|
n\rangle, \label{psitil}
\end{equation}
where
\be \bar{n}=\langle \psi | \hat{n}| \psi \rangle =\sum_{n=1}^{\infty
}n\left| c_{n}\right| ^{2} \label{npsi} \ee
is the mean value of the quantum number operator $\hat{n}\equiv
\hat{a}^{\dagger}\hat{a}$ in $|\psi\rangle$.
(In this article we consider mainly pure states, with the only aim
to simplify notation.)
Designating the mean number of quanta in the state
$\left| \psi_{-}\right\rangle$ as
\be N_{-}=\langle \psi_{-} | \hat{n}| \psi_{-}\rangle
=\frac{1}{\bar{n}}\sum_{n=1}^{\infty }n(n-1)\left| c_{n}\right| ^{2},
\label{npsi-} \ee
it is easy to see that the difference between $N_{-}$ and $\bar{n}$
(the `photon excess') is nothing but Mandel's $q$-parameter
\cite{Man-Q}:
\be
 N_{-} - \bar{n} =\overline{n^2}/\bar{n}\,  -1 -\bar{n}
\equiv q \,. \label{q-Mand} \ee

Note that the transformation $|\psi\rangle \to |\psi_-\rangle$
is not a pure mathematical exercise. It is a real consequence
of conditional measurements on beam splitters \cite{Dakna,split}.
Moreover, according to
the Srinivas--Davies  theory of photodetection \cite{srinivas},
the quantum states of
field modes undergo the same transformation upon detection
of one photon. In the last case,
equation (\ref{q-Mand}) and the related interpretation of
the physical meaning of Mandel's parameter were discussed in \cite{Ueda,Lee}.

As follows from (\ref{q-Mand}),
the `annihilation' operator $\hat{a}$ diminishes
effectively the mean number of quanta ($q <0$) only
for states having {\em sub-Poissonian statistics} (which have been
always considered as `nonclassical').
There are many families of `sub-Poissonian' states, besides the Fock ones.
In particular, all `binomial' states \cite{AhaLer,binom}
possess sub-Poissonian photon statistics. Other examples include
the families of so-called `Barut--Girardello states' \cite{BG,BriBen94}
or odd coherent states \cite{DMM74}.

For any super-Poissonian state,
the action of the `annihilation' operator results in {\em increase} of
the mean number of quanta.
The explanation of such a `paradoxical'
behaviour stems from the first factor, $\sqrt{n}$, in
the right-hand side of equation (\ref{an}): besides
 shifting the Fock state $|n\rangle$ by one
quantum to the left, the operator $\hat{a}$ also increases
the amplitude of the new state. Therefore, under certain
conditions the relative weight of Fock states with larger numbers of
quanta in state (\ref{psitil}) can be greater than in the initial
state (\ref{psiin}), which results finally in the increase of the mean
number of quanta (cf. \cite{Ueda}).

\subsection{Superposition of two Fock states}

Suppose the superposition of two different Fock states
\be |\psi\rangle_{F}=\sqrt{r}\,|n\rangle + \sqrt{1-r}\,|m\rangle,
\qquad 0<r<1. \label{supFock} \ee Then
\[
\bar{n}=r n + (1-r)m, \qquad N_{-} = \frac{r n(n-1) + (1-r)m(m-1)}{r n
+ (1-r)m},
\]
and the condition $N_{-} > \bar{n}$ is equivalent to the inequality
$r(1-r)(n-m)^2 > r(n-m) +m$, which can be satisfied for any $r$
(different from zero and $1$) if the difference $n-m$ is sufficiently
large.

Since there are no limitations on the positive values of Mandel's
parameter, applying the `annihilation' operator to highly
super-Poissonian states one can increase the mean number of quanta by
any desired quantity. For example, in the case of state
(\ref{supFock}) one has $q\approx (1-r)(n-m)$ if $r(n-m)\gg m$, and
one can obtain $q\gg 1$ if also $(1-r)(n-m) \gg 1$. On the other
hand, by calculating the $q$-factor for the state
$|\psi_{-}\rangle_{F}$, one can see that it tends to the limit value
\[
q_{-}^{\infty}=\frac{1-r}{r}\, m -1,
\]
when $n\to\infty$, for fixed $m$ and $r$. So, if $(1-r)m\ll 1$, but
$(1-r)n \gg 1$, then by applying the operator $\hat{a}$ on
$|\psi\rangle$ one can significantly {\em increase \/} the mean number
of quanta and simultaneously transform the  highly
super-Poissonian state $|\psi\rangle$ into sub-Poissonian state $|\psi
_- \rangle$, which turns out to be much more closer to the state
$|n-1\rangle$ than the initial state was, with respect to $|n\rangle$.
This happens due to a significant reduction of the relative weight of
the low-energy state. Indeed, if initially this weight was equal to
$(1-r)/r$, in the new state it becomes approximately $n/m \gg 1$ times
smaller. In other words, the operator $\hat{a}$ effectively
`annihilates' the  {\em low-energy components\/} of quantum superpositions,
{\em increasing the relative weight\/} of the higher-energy components.


\section{`Hyper-Poissonian' states}
Acting on the state $|\psi\rangle$
by the operator $\hat{a}^{\dagger}$
we obtain the `photon-added state' \cite{Dakna,LuH,Wang00,phot-ad}
\begin{equation}
\left| \psi_{+}\right\rangle =
\frac{1}{\sqrt{1+\bar{n}}}\hat{a}^{\dagger} | \psi \rangle =
\frac{1}{\sqrt{1+\bar{n}}} \sum_{n=0}^{\infty }c_{n}\sqrt{1+n}\,|
n+1\rangle . \label{psiplus}
\end{equation}
The mean value of the quantum number operator
in the state $\left| \psi_{+}\right\rangle$ equals
\be N_{+}=\langle
\psi_{+} | \hat{n}| \psi_{+}\rangle
=\frac{1}{1+\bar{n}}\sum_{n=0}^{\infty }(n+1)^2\left| c_{n}\right| ^{2}
= \bar{n} +1 +
\frac{\overline{(\Delta n)^2} }{1+\bar{n}}.
\label{npsi+}
\ee
The difference $N_{+} -\bar{n}$ is always not less than $1$ (being
equal to $1$ only for the Fock states).
Therefore the name `creation operator'
is justified for the operator $\hat{a}^{\dagger}$. Nevertheless,
in certain cases one can add much
more quanta by acting on state $|\psi \rangle$ not with the `creation'
operator, but with the `annihilation' partner!
Using the relation $\overline{(\Delta n)^2}= \bar{n}(1+q)$ together
with equations (\ref{q-Mand}) and (\ref{npsi+}) one can see that the
condition $N_{-} > N_{+}$ is equivalent to
\be
q> 1 + 2\bar{n}.
\label{hyper-cond}
\ee
The states possessing the property $q> \bar{n}$ were named in
\cite{McNeil75,Sot81} `super-chaotic' or `super-random' states
(for such states, the second-order correlation parameter $g^{(2)}(0)$
is greater than its value for the thermal states
$g^{(2)}_{therm}(0)=2$).
Therefore the states possessing the property (\ref{hyper-cond}) can be
named `hyper-chaotic' or `hyper-Poissonian'.
Note that Mandel's parameter of the {\em squeezed vacuum states\/}
is given exactly by the right-hand side of equation (\ref{hyper-cond})
(it can be calculated, e.g., using the results of \cite{1mod}).
Consequently, the `hyper-Poissonian' states can be characterized as those
for which the fluctuations of the photon number are stronger than in the
squeezed vacuum states.

\subsection{Superposition of coherent and vacuum states}

Let us consider a superposition of the {\em coherent state\/}
$|\alpha\rangle$ and the {\em vacuum state\/} $|0\rangle$
\be |\psi\rangle_{0\alpha} =\sqrt{\eta}\,|\alpha\rangle +\xi\,
|0\rangle, \qquad \eta +|\xi|^2
+2\sqrt{\eta}\,e^{-\alpha^2/2}\mbox{Re}\xi=1. \label{psicohvac} \ee
Since the overall phases are not essential, we assume that $\eta$ and
$\alpha$ are real positive numbers (whereas the coefficient $\xi$ may be
complex; however, it does not enter the formulae which we are
interested in). In this case the operator $\hat{a}$ `annihilates' the
vacuum component of the initial superposition (\ref{psicohvac}),
transforming $|\psi\rangle_{0\alpha}$ into the coherent state
$|\psi_{-}\rangle_{0\alpha}=|\alpha\rangle$.
This transformation is accompanied by an increase of the
mean number of quanta under certain conditions:
\be \bar{n}=\eta\alpha^2, \qquad N_{-}= \alpha^2, \qquad
N_{+}=\frac{\eta\alpha^4 + 3\eta\alpha^2 +1}{1 + \eta\alpha^2}\,.
\label{meancohvac} \ee
The ratio $N_{+}/N_{-}$ equals
\[%
\frac{N_{+}}{N_{-}}=1-\frac{1- \alpha^{-2}-3 \eta}{1+\eta\alpha^2},
\]%
thus for $3\eta + \alpha^{-2} < 1$ we have $N_{+}<N_{-}$.
Moreover, if $\alpha\gg 1$, but $\eta\alpha^2 \ll 1$, then
$ \bar{n} \ll 1< N_{+}  \ll  N_{-}$.
The probabilities of finding $n$ quanta in the superposition
(\ref{psicohvac}) are
\be
p_0=1-\eta \left( 1 - e^{-\alpha ^2} \right), \quad p_n= \eta
e^{-\alpha ^2} \frac{\alpha^{2n}}{n!}\, .
\label{probcohvac}
\ee
A special case of the state (\ref{psicohvac}) without vacuum component
($p_0=0$) was considered in \cite{Leenonc} as an example of
`the most nonclassical' state (see also a comment in \cite{Basil}).

\subsection{Generating functions and negative binomial states}
It is well known that the statistical properties of the probability
distribution $\{p_n\equiv |c_n|^2\}$ can be calculated with the aid of
the generating function
\begin{equation}
G(z)=\sum_{n=0}^{\infty }p_{n}z^{n},  \qquad
p_{n}=\frac{1}{n!}G^{(n)}(0), \qquad G(1)=1,
\label{generating}
\end{equation}
since its derivatives at $z=1$ generate the factorial moments,
\begin{equation}
\left.d^{r}G(z)/dz^r\right\vert_{z=1}=\overline{n(n-1)(n-2)...(n-r+1)}
\equiv n^{(r)}. \label{factmom}
\end{equation}
If the function $G(z)$ for the state $|\psi\rangle$ is known, then the
functions $G_{\pm}(z)$ for the states $|\psi_{\pm}\rangle$ are given by
the formulae
\be G_{-}(z)=\frac{1}{\bar{n}}\frac{dG}{dz}, \qquad
G_{+}(z)=\frac{1}{1+\bar{n}}z\frac{d}{dz}[zG(z)]. \label{Gpm} \ee
Applying these relations to the super-Poissonian {\em negative
binomial state\/} \cite{AharLer71,JoLa89}
 \be
|\xi,\mu\rangle=\sum_{n=0}^{\infty}
\left[(1-\xi)^{\mu}\frac{\Gamma(\mu+n)}{\Gamma(\mu)n!}\xi^n\right]^{1/2}
\:|n\rangle , \qquad \mu>0, \quad 0\le\xi<1, \label{negbin} \ee
for which
\[
q={\xi}/{(1-\xi)}, \qquad \bar{n}=\mu{ q}, \qquad
G(z)=\left(\frac{1-\xi}{1-z\xi}\right)^{\mu},
\]
we find
\[
G_{-}(z)=\left(\frac{1-\xi}{1-z\xi}\right)^{\mu+1}, \qquad
G_{+}(z)=z\left(\frac{1-\xi}{1-z\xi}\right)^{\mu+1}
\frac{1+z\xi(\mu-1)}{1+\xi(\mu-1)}.
\]
Consequently, applying the operator $\hat{a}$ to the state
(\ref{negbin}) several times, one can add each time equal portions of
quanta $q$ ($N_- =\mu q +q = \bar{n} + q$) without changing Mandel's
parameter ($q_- =q $). On the other hand, after applying the operator
$\hat{a}^{\dagger}$ on $|\psi> $ the mean
number of quanta is increased,
\[
N_{+} - \bar{n}= 1 + \frac{\xi}{1-\xi} +
\frac{\xi(\mu-1)}{1+\xi(\mu-1)},
\]
and one can easily verify that for $\mu<1/2<\xi(1-\mu)$ the action of
the `annihilation' operator adds more quanta than the action of the
`creation' operator.

\section{Quantum states with given mean photon number and Mandel's
parameter}

Suppose that we want to find a probability distribution
$\{p_n\}$ with two given parameters: the mean number
$\bar{n}$ and the `mean photon multiplication factor' $\gamma\equiv
N_{-}/\bar{n}$. Since $N_{-}=\left.dG_{-}/dz\right\vert_{z=1}$, we can
rewrite the second condition, using the first of the relations in equation
(\ref{Gpm}), as  $G^{\prime\prime}(1)=\gamma\bar{n}G^{\prime}(1)$.
Assuming that this relation must hold not only for $z=1$, but also in
some interval of values of the auxiliary variable $z$, including the
point $z=1$, we obtain a simple equation,
$G^{\prime\prime}(z)=\gamma\bar{n}G^{\prime}(z)$, whose solution,
satisfying the conditions $G(1)=1$ and $G^{\prime }(1)=\bar{n}$, reads
\be G(z)= 1-\,\frac{1}{\gamma} +
\,\frac{1}{\gamma}e^{\gamma\bar{n}(z-1)}. \label{Ggamma} \ee The
corresponding `photon number' distribution is \be p_0=
1-\,\frac{1}{\gamma} + \,\frac{1}{\gamma}e^{-\gamma\bar{n}}, \qquad
p_n= \frac{1}{\gamma}e^{-\gamma\bar{n}}\, \frac{(\gamma\bar{n})^n}{n!},
\quad n\ge 1. \label{pngam} \ee

A {\em pure} quantum state possessing the photon number
probability distribution $\{p_n\}$ can be written as
\be |\psi\rangle = \sum_{n=0}^{\infty}
e^{i\phi_n}\sqrt{p_n}\,|n\rangle, \label{psipn} \ee
where  the set of phases $\{\phi_n\}$ may be quite arbitrary. In
particular, one can adjust the phases in such a way that function
(\ref{psipn}) together with the weight set (\ref{pngam}) become
a special case of the superposition (\ref{psicohvac}), with
$|\alpha|^2=\gamma\bar{n}$ and $\eta= \gamma^{-1}$.
The probability distribution (\ref{pngam}) exists for any value of
$\bar{n}$ if $\gamma \ge 1$. For $\gamma <1$ (sub-Poissonian case) it
has sense only for not very large values of the initial mean photon
number: $\gamma\bar{n} \le |\ln(1-\gamma)|$. For $\gamma \to 1$
the Poisson distribution is recovered, while for $\gamma \to \infty$,
$p_0 =1$ and all other probabilities go to zero.

Of course, there are infinitely many different distributions resulting
in the same values of two parameters, $\bar{n}$ and $\gamma$ or
$\bar{n}$ and $q=(\gamma-1)\bar{n}$. For example, one can start from
the equation $G^{\prime\prime}(z)=\gamma\bar{n}^2G(z)$, which results
in the function \be G(z)=\cosh\left[\bar{n}\sqrt{\gamma}\,(z-1)\right]
+\gamma^{-1/2} \sinh\left[\bar{n}\sqrt{\gamma}\,(z-1)\right].
\label{badfun} \ee However, this function can be interpreted as
generating function for nonnegative probabilities $\{p_n\}$ only under
certain restrictions on the values of $\bar{n}$ and $\gamma$, because
the coefficient $p_1$ in the Taylor expansion (\ref{generating}) of
function (\ref{badfun}) becomes negative for $\bar{n}\sqrt{\gamma} \gg
1$ and $\gamma >1$, whereas for $p_0$ the same happens if
$\bar{n}\sqrt{\gamma} \gg 1$ and $\gamma <1$. This example shows the
main difficulty of finding new distributions through differential
equations for their generating functions: although one can write
infinitely many differential equations for $G(z)$ which result in the
necessary conditions at $z=1$, only few of them can be solved
analytically, and even much fewer functions thus obtained have  {\em
all} derivatives at $z=0$ nonnegative. Nonetheless, several
interesting distributions can be found in this way.

\subsection{Generalized logarithmic states}
For example, looking for probabilities resulting in a given value of
the `photon excess' (or Mandel's parameter) $q$, we can rewrite
equation (\ref{q-Mand}) (or $\overline{n^2} - \overline{n}^2
-(q+1)\overline{n}=0$) as (using relations (\ref{factmom}))
\begin{equation}
G^{\prime \prime }(1)-\left[ G^{\prime }(1)\right] ^{2}-qG^{\prime
}(1)=0. \label{eqn2}
\end{equation}
One of many possible extensions of this equality to a finite interval
of values of variable $z$ is the equation
\begin{equation}
G^{\prime \prime }(z)-\left[ G^{\prime }(z)\right] ^{2}-qG^{\prime
}(z)=0 . \label{eqn3}
\end{equation}
Although being nonlinear this equation is easily solved by several
methods, for example, multiplying it on the left by the integrating
factor $e^{-G}$  one gets a linear equation ${dF}/{dz}=qF$ for
$F=e^{-G}G^{\prime }$. Taking into account the conditions $G(1)=1$ and
$G^{\prime }(1)=\bar{n}$, we obtain
\begin{equation}
G(z)=1-\ln \left[ 1+\frac{\bar{n}}{q}\left( 1-e^{q\left( z-1\right)
}\right) \right] . \label{geratriz1}
\end{equation}
The probability distribution for $q>0$ is
\begin{equation}
p_{0}=\ln \left[\frac{qe}{q +\bar{n}\left( 1-e^{-q}\right)} \right],
\qquad p_{n}=\frac{q^{n}}{n!}\sum_{k=1}^{\infty }k^{n-1} \left(
\frac{\bar{n}e^{-q}}{q+\bar{n}}\right) ^{k},\quad n \ge 1. \label{pn1}
\end{equation}
Although all coefficients $p_n$ with $n\ge 1$ are positive for any
positive $q$ and $\bar{n}$, the vacuum probability $p_0$ is
nonnegative provided
\begin{equation}
\bar{n}\le \frac{q\left( e-1\right) }{1-e^{-q}}. \label{cond}
\end{equation}
For $q\gg 1$ and $q\gg \bar{n}$ the probabilities
(\ref{pn1}) become close to (\ref{pngam}):
\begin{equation}
p_{0}\approx 1-\bar{n}/q,\qquad p_{n}\approx \frac{\bar{n}}{q}\left( \frac{%
q^{n}}{n!}e^{-q}\right) ,\quad n=1,2,3,...\,. \label{p04}
\end{equation}

Using equations (\ref{factmom}) and (\ref{Gpm}) we obtain Mandel's
parameter $q_{-}$ for the state $|\psi_{-}\rangle$ (\ref{psitil}) in
terms of the factorial moments of the initial arbitrary state
$|\psi\rangle$
\be q_{-}=\frac{\bar{n}n^{(3)} - [n^{(2)}]^2}{\bar{n}n^{(2)}}.
\label{q-gen} \ee
The peculiar feature of the state generated by function
(\ref{geratriz1}) is that for this state $q_{-}=\bar{n}$, i.e., while
for state $|\psi\rangle $,
$\bar{n}$ is the mean photon number and $q
$ is Mandel's parameter, for state $|\psi_{-}\rangle$ the mean photon
number is $\bar{n}+q$ and Mandel's parameter is $q_{-}=\bar{n}$.
And the `excess of photons' of state $\hat{a}|\psi _- \rangle$ is $\bar{n}$.

Going to the limit $q\to 0$ in (\ref{geratriz1}) we obtain the
generating function
\begin{equation}
G(z)=1-\ln \left[ 1-\bar{n}( z-1) \right], \label{G0}
\end{equation}
which yields the following probabilities and factorial moments
(\ref{factmom}):
\be p_{0} =1-\ln \left( 1+\bar{n}\right) , \qquad p_{n}
=\frac{1}{n}\left( \frac{\bar{n}}{1+\bar{n}}\right) ^{n}, \quad n\ge
1, \label{prob0} \ee
 \be n^{(r)} =(r-1)!\;\bar{n}^r, \qquad
r=1,2,\ldots\,. \label{facmom0} \ee
The probability distribution with the same factorial moments
(\ref{facmom0}) was found in \cite{Bal79} practically by the same
method. The difference is that the authors of \cite{Bal79} calculated
the generating function not for the probabilities, but for the
factorial moments, having obtained, in particular, the function
\be Q_1(x)\equiv \sum_{\nu=0}^{\infty} (-x)^{\nu} \langle
\hat{a}^{\dagger\nu}\hat{a}^{\nu}\rangle/\nu!
=(A-1)^{-1}\ln\left[(1-A)\bar{n}x+1\right] +1, \label{Q1Bal}
\ee
where the {\em positive} constant $A$ was considered as a measure of
the relative deviation from the Poissonian case (the article
\cite{Bal79} appeared before \cite{Man-Q}):
\be \left[\langle \hat{a}^{\dagger}\hat{a}\rangle
(\hat{a}^{\dagger}\hat{a}\rangle-1)\rangle -\langle
\hat{a}^{\dagger}\hat{a}\rangle^2\right]/ \langle
\hat{a}^{\dagger}\hat{a}\rangle^2 = -A. \label{def-A} \ee
In the limit case $A=0$ formula (\ref{Q1Bal}) yields the factorial
moments in the form (\ref{facmom0}) (but for $A\neq 0$ and
$q=-A\bar{n}\neq 0$ the functions (\ref{geratriz1}) and (\ref{Q1Bal})
give different values of factorial moments).

The statistics of the probability distribution (\ref{prob0}) (not found
explicitly in \cite{Bal79}) is quite different from the Poissonian
statistics, although this distribution gives Mandel's factor $q=0$
and the same first two factorial moments as in the coherent state with
$|\alpha|^2=\bar{n}$. The corresponding quantum state was named in
\cite{Bal79} {\em sub-coherent state} (the same name was used in
\cite{BriBen94} for quite different states possessing sub-Poissonian
statistics).
We can note in this connection
that one of many possible pure states corresponding to the
distribution (\ref{prob0}) is a special case of the family of
so-called {\em logarithmic states} introduced in \cite{Simon}
\be |\psi\rangle_{log}= c|0\rangle +\sum_{n=1}^{\infty}
\frac{z^n}{\sqrt{n}} \:|n\rangle, \quad
|c|^2=1+\ln\left(1-|z|^2\right), \quad
\overline{n}=\frac{|z|^2}{1\!-\!|z|^2}. \label{log} \ee

\section{A distinguished role of the exponential phase operator
and phase coherent states}

Acting on the `logarithmic' state (\ref{log}) by the operator
$\hat{a}$ we arrive at its `subtracted' partner
\be |\psi_{-}\rangle_{log}= \sqrt{1\!-\!|z|^2} \sum_{n=0}^{\infty} z^n
\:|n\rangle, \label{phase} \ee
known under the names {\em coherent phase state\/} \cite{Ler,VBM96,Wun01},
{\em harmonious state} \cite{Sudar93}, or {\em pseudothermal state}
\cite{DoMi} (see also \cite{rev-ncs}).

The mean number of quanta in the `sub-coherent' state (\ref{log})
cannot exceed a small value $e-1 \approx 1.72$. However, it is easy to
show that there exist many {\em sub-coherent states} ($q=0$) with
arbitrary values of $\bar{n}$. An example is the superposition of two
Fock states (\ref{supFock}). Considering for simplicity the case of
$n\gg m$, one can verify that $q=0$ either for $r\approx m/n^2$ (when
$\bar{n}\approx m$) or for $1-r\approx 1/n$ (when $\bar{n}\approx n$).

The coherent phase state (\ref{phase}) is the eigenstate of the {\em
exponential phase operator}
\be \hat{E}_-\equiv \sum_{n=1}^\infty|n-1\rangle \langle n|
=\left(\hat{a}\hat{a}^{\dagger}\right)^{-1/2}\hat{a} \equiv
\left(\hat{n} +1\right)^{-1/2}\hat{a}, \label{E-} \ee
introduced in \cite{Suss} and discussed in \cite{VBM96,Wun01,CarNiet}.
Its Hermitian conjugated partner is
\be \hat{E}_+ = \sum_{n=1}^\infty|n\rangle \langle n-1| =
\hat{a}^{\dagger}\left(\hat{a}\hat{a}^{\dagger}\right)^{-1/2} \equiv
\hat{a}^{\dagger} \left(\hat{n} +1\right)^{-1/2}, \label{E+} \ee
\[%
[\hat{E}_- ,\hat{E}_+]=|0\rangle \langle 0|, \qquad \hat{E}_+
\hat{E}_- =\hat{1}- |0\rangle \langle 0|.
\]%
Acting on the Fock states, the operators $\hat{E}_{\pm}$ shift the
number of quanta by $\pm 1$ {\em without changing the amplitude} of
the state vector:
\be%
\hat{E}_-|n\rangle =(1-\delta_{n0})|n-1\rangle, \quad
\hat{E}_+|n\rangle =|n+1\rangle. \label{actEpm}
\ee%
Therefore, applying the operators $\hat{E}_{\pm}$ to an arbitrary state
$|\psi\rangle$ (\ref{psiin}) we obtain the following normalized states
$|\widetilde\psi_{\pm}\rangle \sim \hat{E}_{\pm}|\psi\rangle$:
\be |\widetilde\psi_{-} \rangle =\frac{1}{\sqrt{1-|c_0|^2}}
\sum_{n=0}^{\infty }c_{n+1}| n\rangle, \qquad |\widetilde\psi_{+}
\rangle = \sum_{n=0}^{\infty }c_{n}| n+1\rangle. \label{psitilpm}
\ee
Their quantum number generating functions are related to the initial
generating function $G(z)$ (\ref{generating}) by simple relations
\be \widetilde{G}_{-}(z)=\frac{G(z)-G(0)}{z(1-p_0)}, \qquad
\widetilde{G}_{+}(z)= zG(z). \label{tilGpm}
\ee
Consequently, the mean numbers of quanta in the new states
(\ref{psitilpm}) are connected with the mean number $\bar{n}$ in the
initial state as follows:
\be \widetilde{N}_{-}\equiv\langle \widetilde\psi_{-} | \hat{n}|
\widetilde\psi_{-}\rangle = \frac{\bar{n}}{1-p_0} - 1, \qquad
\widetilde{N}_{+}\equiv\langle \widetilde\psi_{+} | \hat{n}|
\widetilde\psi_{+}\rangle = \bar{n} + 1. \label{tilmeanpm}
\ee
We see that the operator $\hat{E}_+$ adds {\em exactly one} quantum
to {\em any} quantum state, whereas the operator $\hat{E}_-$ {\em
removes} exactly one quantum from any state {\em which has no
contribution of the vacuum state} (i.e., if $p_0=0$). Therefore,
perhaps, not operators $\hat{a}$ and $\hat{a}^{\dagger}$, but
operators $\hat{E}_-$ and $\hat{E}_+$ should be named `annihilation'
and `creation' operators in the literal meaning of these words!
(See in this connection study \cite{Moya99}, where it was shown how
the states $\hat{E}_+^m|\psi\rangle$ could arise as a result of the
interaction of the Jaynes--Cummings type between atoms and cavity
fields or between motional and internal degrees of freedom of trapped
ions.
{\em Shifted thermal states\/}, which can be written as
$\hat\rho_{th}^{(shift)}=\hat{E}_+^{m}\hat\rho_{th}\hat{E}_-^{m}$,
have been considered in \cite{Lee97},
whereas methods of
generating such states in a micromaser were discussed in \cite{SMW96}.)

For the coherent states $p_0=\exp(-\bar{n})$, and for $\bar{n}\ll 1$ we have
$\widetilde{N}_{-}=\bar{n}/2$. For the coherent phase states (\ref{phase})
we have $\widetilde{N}_{-} \equiv\bar{n}$ for any
$\bar{n}=|z|^2/\left(1-|z|^2\right)$.
For many other states with $p_0>0$, one can always make
the `modified photon excess'
\be \widetilde{q} \equiv \widetilde{N}_{-} -\bar{n}=
\frac{p_0 \bar{n}}{1-p_0} - 1
\label{tilq} \ee
positive by increasing the value of $\bar{n}$,
which is, in a generic case, a free parameter independent of $p_0$.
This can be easily
seen, e.g., in the example (\ref{pngam}), when the quantity
\[
(1-p_0)\widetilde{q}= \bar{n} -\, \frac{\bar{n}+1}{\gamma}
\left(1-e^{-\gamma\bar{n}}\right)
\]
is obviously positive for sufficiently large $\bar{n}$ and $\gamma>1$.
The same is true for the negative binomial states (\ref{negbin}) with
$\mu<1$ and $\xi$ sufficiently close to $1$.
Moreover, if $\bar{n}>2\left(1-p_0\right)/p_0$, then
$\widetilde{N}_{-} > \widetilde{N}_{+}$.

Even using operators of the form
$\hat{A}=f(\hat{n})\hat{a}$ we can find the states $|\psi\rangle$ for
which state $\hat{A}|\psi\rangle$ has more quanta (in average) than
the state $|\psi\rangle$, for {\em arbitrary function\/}
$f(\hat{n})$. The simplest example is the `two-Fock' state
(\ref{supFock}) with $m=0$. In this case $\bar{n}=rn$. Any operator
$f(\hat{n})\hat{a}$ transforms this superposition into the single
Fock state $|n-1\rangle$, and obviously $(n-1)>rn$ if $(1-r)>1/n$.

\section{Conclusion}
Concluding, we have shown that the standard `annihilation' boson
operator diminishes the mean number of quanta only for `nonclassical'
states with negative values of Mandel's parameter, whereas there are
many  states (all of them {\em superpositions} or mixtures of the Fock
states), with the peculiarity that when acted by $\hat{a}$ produce new
states whose mean number of quanta is increased. Moreover, we have
shown that in certain cases the `annihilation' operator $\hat{a}$ can
increase the mean number of quanta much more effectively than its
`creation' partner $\hat{a}^{\dagger}$.
We emphasized an interpretation (discovered in \cite{Ueda,Lee}
but still not well known)
of Mandel's $q$-factor as the `photon
excess' in the mean number of quanta arising as a result of action of
operator $\hat{a}$ on the given quantum state.
 Besides, we have shown that maybe the pair of Hermitian
conjugated {\em exponential phase operators} could be the best
candidates for being  called `annihilation-creation' operators,
although the `lowering' operator from this pair also cannot diminish
the mean number of quanta by exactly one unit, for quantum states with
nonzero vacuum component.
This gives a hint that, perhaps, in the  quantum
photocount formalism of continuous measurements, as for instance, that
proposed by Srinivas and Davies \cite{srinivas} one should use
instead of $\hat{a}$ and $\hat{a}^{\dagger}$
the operators $\hat{E}_-$ and $\hat{E}_+$,
which would not violate the property V of the cited
work (namely, the condition of boundness for the counting rate).
However, this issue is outside the objective of this letter, and
we shall report on it elsewhere.

{SSM acknowledges partial financial support by CNPq (Bras\'{\i}lia,
DF) and FAPESP (S\~ao Paulo; Contract No. 2000/15084-5). VVD
thanks CNPq for the full financial support.}



\begin{thebibliography}{99}

\bibitem{annih}
Bohm A 1994 {\em Quantum Mechanics: Foundations and Applications\/}
(Berlin: Springer) p 53\\
Mandel L and Wolf E 1995 {\em Optical Coherence and Quantum Optics\/}
(Cambridge: Cambridge University Press) p 479\\
Meystre P and Sargent III M 1991 {\em Elements of Quantum Optics\/}
(Berlin: Springer) p 94\\
Scully M O and Zubairy M S 1997 {\em Quantum Optics\/}
(Cambridge: Cambridge University Press) p 5\\
Walls D F and Milburn G J 1994 {\em Quantum Optics\/} (Berlin:
Springer) p 11

\bibitem{destr}
Bjorken J D and Drell S D 1965 {\em Relativistic Quantum Fields\/}
(London: McGraw Hill) p 39\\
Cohen-Tannoudji C, Diu B and Lalo\"e F 1977 {\em Quantum Mechanics\/}
(New York: Wiley) p 494\\
Loudon R 1994 {\em The Quantum Theory of Light\/}
(Oxford: Clarendon) p 129\\
Schiff L I 1968 {\em Quantum Mechanics\/} (London: McGraw Hill) p 183

\bibitem{sakurai} Sakurai J J 1967 {\em Advanced Quantum Mechanics}
(Reading: Addison Wesley) in p 27 attributes gods names, from
the Hindu mythology, to the operators: $\hat{a}$ is {\bf Siva}, the
Destroyer, $\hat{a}^{\dagger}$ is {\bf Brahma}, the Creator and $
\hat{n}= \hat{a}^{\dagger}\hat{a}$ is {\bf Vishnu}, the Preserver.

\bibitem{Dirac} Dirac P A M 1987 {\em The Principles of Quantum Mechanics\/}
4th ed (Oxford: Oxford University Press)

\bibitem{Fock} Fock V 1928
{\em Z. Phys.}  {\bf 49} 339

\bibitem{Dakna}
Dakna M, Kn\"oll L and Welsch D-G 1998 {\em Europ. Phys. J.} D  {\bf 3} 295

\bibitem{LuH} Hong L 1999
{\em Phys. Lett.} A {\bf 264} 265

\bibitem{Wang00} Wang X G 2000
{\em Opt. Commun.}  {\bf 178} 365

\bibitem{Man-Q} Mandel L 1979
{\em Opt. Lett.} {\bf 4} 205

\bibitem{split} Mattle K, Michler M, Weinfurter H, Zeilinger A
and Zukowski M 1995
{\em Appl. Phys.} B {\bf 60} S111 \\
%
Ban M 1996 {\em J. Mod. Opt.}  {\bf 43} 1281 \\
%
Lvovsky A I and Mlynek J 2002
{\em Phys. Rev. Lett.} {\bf 88} 250401

\bibitem{srinivas} Srinivas M D and Davies E B 1981 {\em Optica Acta\/}
{\bf 28} 981

\bibitem{Ueda} Ueda M, Imoto N and Ogawa T 1990
{\em Phys. Rev.} A  {\bf 41} 3891

\bibitem{Lee} Lee C T 1993 {\em Phys. Rev.} A  {\bf 48} 2285

\bibitem{AhaLer}
Aharonov Y, Lerner E C, Huang H W and Knight J M {\em J. Math. Phys.}
1973 {\bf 14} 746

\bibitem{binom}  Stoler D,  Saleh B E A and  Teich M C 1985
{\em Opt. Acta\/} {\bf 32} 345 \\
%
Lee C T 1985
{\em Phys. Rev.} A {\bf 31} 1213

\bibitem{BG} Barut A O and Girardello L 1971
{\em Commun. Math. Phys.} {\bf 21} 41

\bibitem{BriBen94} Brif C and Ben-Aryeh Y 1994
{\em Quant. Opt.} {\bf 6} 391

\bibitem{DMM74} Dodonov V V,  Malkin I A and  Man'ko V I 1974
{\em Physica\/} {\bf 72} 597


\bibitem{phot-ad}  Agarwal G S and  Tara K  1991
{\em Phys. Rev.} A  {\bf 43} 492 \\
%
Zhang Z and  Fan H 1992
{\em Phys. Lett.} A  {\bf 165} 14  \\
%
Kis Z, Adam P and Janszky J 1994
{\em Phys. Lett.} A  {\bf 188} 16 \\
%
%
Xin Z Z, Duan Y B, Zhang H M, Hirayama M and  Matumoto K I 1996
{\em J. Phys.} B  {\bf 29} 4493  \\
%
Dodonov V V,  Korennoy Y A,  Man'ko V I and Moukhin Y A  1996
{\em Quant. Semicl. Opt.}  {\bf 8} 413 \\
Man'ko V I and W\"{u}nsche A  1997 {\em Quant. Semiclass. Opt.} {\bf 9} 381 \\
%
%
Sixdeniers J M and Penson K A  2001
{\em J. Phys. A: Math. Gen.}   {\bf 34} 2859 \\
%
Quesne C 2001 {\em Phys. Lett.} A {\bf 288} 241

\bibitem{McNeil75} McNeil K J and Walls D F 1975
{\em Phys. Lett.} A {\bf 51} 233

\bibitem{Sot81} Sotskii B A and Glazachev B I 1981
{\em Opt. Spectrosc.} {\bf 50} 582

\bibitem{1mod} Dodonov V V, Man'ko O V and Man'ko V I 1994
{\em Phys. Rev.} A {\bf 49} 2993

\bibitem{Leenonc} Lee C T 1995 {\em Phys. Rev.} A  {\bf 52} 3374

\bibitem{Basil} de Lima A F and Baseia B 1996
{\em Phys. Rev.} A  {\bf 54} 4589

\bibitem{AharLer71} Aharonov Y, Huang H W, Knight J M and
Lerner E C
{\em Lett. Nuovo Cim.} 1971 {\bf 2} 1317

\bibitem{JoLa89}  Joshi A and Lawande S V 1989
{\it Opt. Commun.}  {\bf 70} 21 \\
%
Matsuo K 1990 {\it Phys. Rev.} A   {\bf 41} 519


\bibitem{Bal79} Baltes HP, Quattropani A and Schwendimann P  1979
{\em J. Phys. A: Math. Gen.} {\bf 12} L35

\bibitem{Simon} Simon R and Satyanarayana M V 1988
{\it J. Mod. Opt.} {\bf 35} 719

\bibitem{Ler} Lerner E C, Huang H W and Walters G E 1970
{\em J. Math. Phys.} {\bf 11} 1679 \\
%
Ifantis E K 1972 {\em J. Math. Phys.}  {\bf 13} 568 \\
%
Shapiro J H and Shepard S R 1991
{\em Phys. Rev.} A {\bf 43} 3795 \\
%
Chaturvedi S, Kapoor A K, Sandhya R, Srinivasan V and Simon R 1991
{\em Phys. Rev.} A {\bf 43} 4555 \\
%
Vourdas A 1992
{\em Phys. Rev.} A {\bf 45} 1943 \\
%
Hall M J W 1993  {\em J. Mod. Opt.} {\bf 40} 809 \\
Brif C and Ben-Aryeh Y 1994
{\em Phys. Rev.} A  {\bf 50} 3505

\bibitem{VBM96} Vourdas A, Brif C and Mann A 1996
{\em J. Phys. A: Math. Gen.} {\bf 29} 5887

\bibitem{Wun01}
W\"unsche A 2001 {\em J. Opt.} B {\bf 3} 206


\bibitem{Sudar93} Sudarshan E C G 1993
{\em Int. J. Theor. Phys.}  {\bf 32} 1069

\bibitem{DoMi} Dodonov V V and Mizrahi S S 1995
{\em Ann. Phys.} (NY) {\bf 237} 226

\bibitem{rev-ncs} Dodonov V V 2002
{\em J. Opt.} B {\bf 4} R1

\bibitem{Suss} Susskind L and Glogower J  1964
{\em Physics\/}  {\bf 1} 49

\bibitem{CarNiet} Carruthers P and Nieto M 1968
{\em Rev. Mod. Phys.}  {\bf 40} 411 \\
%
Paul H 1974
{\em Fortschr. Phys.}  {\bf 22} 657  \\
%
Bergou J and Englert B-G  1991
{\em Ann. Phys.} (NY)  {\bf 209} 479 \\
%
Luk\v s A and Pe\v rinov\'a V 1994
{\em Quant. Opt.}  {\bf 6} 125  \\
%
Ban M 1995 {\em Phys. Lett.} A {\bf 199} 275 \\
Lynch R 1995 
{\em Phys. Rep.}  {\bf 256} 367 \\
Royer A 1996
{\em Phys. Rev.} A {\bf 53} 70

\bibitem{Moya99} Moya-Cessa H, Chavez-Cerda S and Vogel W 1999
{\em J. Mod. Opt.} {\bf 46} 1641

\bibitem{Lee97} Lee C T 1997
{\em Phys. Rev.} A  {\bf 55} 4449

\bibitem{SMW96} Scully M O, Meyer G M and Walther H 1996
{\em Phys. Rev. Lett.} {\bf 76} 4144

\end{thebibliography}
\end{document}